\documentclass[letterpaper,titlepage,10pt]{article}
\usepackage[utf8]{inputenc} 
\usepackage{textcomp} 
\usepackage{graphicx}  
\usepackage{flafter}  

\usepackage{amsmath,amssymb}  
\usepackage{bm}  
\usepackage{wasysym}

\usepackage{memhfixc}  
\usepackage{pdfsync}  

\begin{document}
\title{Navier-Stokes regularity in 3D}   
\begin{abstract}
This short proof shows that for smooth and sufficiently fast decaying initial data at infinity, the full incompressible Navier-Stokes solutions are eternal. The proof uses an orthogonal decomposition of the velocity field and some well-known vector calculus identities to establish a particular contradiction, which leads to a vanishing integral, which is the main integral that determines the evolution of enstrophy. As it is shown that enstrophy is non-increasing, it is well-know that the solutions stay regular at all times. \end{abstract}

\author{Jussi Lindgren}
\maketitle

\section{Proof of regularity in $\mathbb{R}^3$}
The full Navier-Stokes equations can be stated as

\begin{equation} \frac{\partial \mathbf{u}}{\partial t}+ \mathbf{u}\cdot \nabla \mathbf{u}=-\nabla p +\Delta \mathbf{u}\end{equation}

with the usual incompressibility condition

\begin{equation}  \nabla \cdot \mathbf{u}=0\end{equation}

We transform the equation in a more useful form using the following vector calculus identity

\begin{equation}  \mathbf{u}\cdot \nabla \mathbf{u}=\frac{1}{2}\nabla (\mathbf{u}\cdot \mathbf{u})-\mathbf{u}\times \mathbf{\omega}\end{equation}

Substituting this back into the Navier-Stokes equation one has

\begin{equation}  \frac{\partial \mathbf{u}}{\partial t}=-\frac{1}{2}\nabla (\mathbf{u}\cdot \mathbf{u})+\mathbf{u}\times \mathbf{\omega}-\nabla p +\nu \Delta \mathbf{u}\end{equation}

Operating with the curl operator, one gets the vorticity equation

\begin{equation}  \frac{\partial \mathbf{\omega}}{\partial t}=\nabla \times (\mathbf{u}\times \mathbf{\omega}) +\nu \Delta \mathbf{\omega}\end{equation}

\subsection{Enstrophy identity in $\mathbb{R}^3$}

Now, first of all we use the divergence theorem. For two vector fields $\mathbf{u}$ and $\mathbf{\omega}$ we have the following result applying the divergence theorem:
\begin{equation}
\iiint_{V}\mathbf{\omega}\cdot (\nabla \times \mathbf{u})-\mathbf{u}\cdot (\nabla \times \mathbf{\omega})dV=\oiint \mathbf{u}\times \mathbf{\omega} \cdot d\mathbf{S}
\end{equation}

Now assuming the velocity field decays fast in infinity, we have for the whole space

\begin{equation}
\iiint_{\mathbb{R}^3}\mathbf{\omega}\cdot (\nabla \times \mathbf{u})dV=\iiint_{\mathbb{R}^3}\mathbf{u}\cdot (\nabla \times \mathbf{\omega})dV\end{equation}

The left hand side is just the enstrophy of the flow, so if we define

\begin{equation}
E(t)=\iiint_{\mathbb{R}^3}\mathbf{\omega}\cdot (\nabla \times \mathbf{u})dV
\end{equation}

we have

\begin{equation}
E(t)=\iiint_{\mathbb{R}^3}\mathbf{u}\cdot (\nabla \times \mathbf{\omega})dV
\end{equation}

We can decompose the velocity field into two parts: into one that is parallel to $\nabla \times \mathbf{\omega}$ and into one which is perpendicular to it. In other words, we have the decomposition
\begin{equation}
\mathbf{u}=\mathbf{u}_{\parallel} +\mathbf{u}_{\perp}
\end{equation}
\begin{equation}
E(t)=\iiint_{\mathbb{R}^3}(\mathbf{u}_{\parallel}+\mathbf{u}_{\perp})\cdot (\nabla \times \mathbf{\omega})dV
\end{equation}

By the defining property of the dot product, we then have

\begin{equation}
E(t)=\iiint_{\mathbb{R}^3}(\mathbf{u}_{\parallel})\cdot (\nabla \times \mathbf{\omega})dV
\end{equation}

Taking time derivative of the enstrophy

\begin{equation}
\frac{d}{dt}E(t)=\frac{d}{dt}\iiint_{\mathbb{R}^3}(\mathbf{u}_{\parallel})\cdot (\nabla \times \mathbf{\omega})dV
\end{equation}

Now we can see that the time derivative of enstrophy does not depend on the velocity component which is perpendicular to $\nabla \times \mathbf{\omega}$. 

\subsection{Differential equation for enstrophy from the vorticity equation and the contradiction}

In order to ultimately obtain the enstrophy equation from the vorticity equation, one needs to dot the vorticity equation with vorticity:

\begin{equation}  \frac{1}{2}\frac{\partial \mathbf{\omega}\cdot \mathbf{\omega}}{\partial t}=\mathbf{\omega}\cdot \nabla \times (\mathbf{u}\times \mathbf{\omega})+\mathbf{\omega}\cdot \nu \Delta \mathbf{\omega}\end{equation}

Now we use the following vector calculus identity:

\begin{equation}  \nabla \cdot (\mathbf{A}\times \mathbf{B})=\mathbf{B}\cdot (\nabla \times \mathbf{A})-\mathbf{A}\cdot (\nabla \times \mathbf{B})\end{equation}

Let us make then the following identification:

\begin{equation}  \mathbf{B}=\mathbf{\omega}\end{equation}

and

\begin{equation}  \mathbf{A}=\mathbf{u}\times \mathbf{\omega}\end{equation}

We then have

\begin{equation}  \mathbf{\omega}\cdot \nabla \times (\mathbf{u}\times \mathbf{\omega})=\nabla \cdot ((\mathbf{u}\times \mathbf{\omega})\times \mathbf{\omega})+(\mathbf{u}\times \mathbf{\omega})\cdot (\nabla \times \mathbf{\omega})\end{equation}

We substitute this expression back  to the enstrophy equation to get:

\begin{equation} \frac{1}{2}\frac{\partial \mathbf{\omega}\cdot \mathbf{\omega}}{\partial t}=\nabla \cdot ((\mathbf{u}\times \mathbf{\omega})\times \mathbf{\omega})+(\mathbf{u}\times \mathbf{\omega})\cdot (\nabla \times \mathbf{\omega})+\mathbf{\omega}\cdot \nu \Delta \mathbf{\omega}\end{equation}

The only problematic term now is the second one on the right side of the equation. Let us consider it more closely:

\begin{equation}  (\mathbf{u}\times \mathbf{\omega})\cdot (\nabla \times \mathbf{\omega})\end{equation}

 Moreover, using scalar triple product, we can write
 \begin{equation}
 (\mathbf{u}\times \mathbf{\omega})\cdot (\nabla \times \mathbf{\omega})=(\nabla \times \mathbf{\omega})\cdot (\mathbf{u}\times \mathbf{\omega})=\mathbf{\omega}\cdot ((\nabla \times \mathbf{\omega})\times \mathbf{u})
 \end{equation}

We then have
\begin{equation}
 (\mathbf{u}\times \mathbf{\omega})\cdot (\nabla \times \mathbf{\omega})=\mathbf{\omega}\cdot ((\nabla \times \mathbf{\omega})\times ( \mathbf{u}_{\parallel}+\mathbf{u}_{\perp}))
\end{equation}

Using the basic property of cross product we then have

\begin{equation}
 (\mathbf{u}\times \mathbf{\omega})\cdot (\nabla \times \mathbf{\omega})=\mathbf{\omega}\cdot ((\nabla \times \mathbf{\omega})\times \mathbf{u}_{\perp})
\end{equation}

Integrating over the whole space:

\begin{equation} \frac{1}{2}\iiint_{\mathbb{R}^3}\frac{\partial \mathbf{\omega}\cdot \mathbf{\omega}}{\partial t}dV=\iiint_{\mathbb{R}^3}\nabla \cdot ((\mathbf{u}\times \mathbf{\omega})\times \mathbf{\omega})dV+\iiint_{\mathbb{R}^3}(\mathbf{u}\times \mathbf{\omega})\cdot (\nabla \times \mathbf{\omega})dV+\iiint_{\mathbb{R}^3}\mathbf{\omega}\cdot \nu \Delta \mathbf{\omega}dV\end{equation}

Substituting and noting that the divergence term vanishes we have

 \begin{equation} \frac{1}{2}\iiint_{\mathbb{R}^3}\frac{\partial \mathbf{\omega}\cdot \mathbf{\omega}}{\partial t}dV=\frac{1}{2}\frac{dE(t)}{dt}=\iiint_{\mathbb{R}^3}\mathbf{\omega}\cdot ((\nabla \times \mathbf{\omega})\times \mathbf{u}_{\perp})dV+\iiint_{\mathbb{R}^3}\mathbf{\omega}\cdot \nu \Delta \mathbf{\omega}dV\end{equation}

Now it seems that the enstrophy here depends on the orthogonal part of the velocity field. This cannot be the case and is a clear contradiction, therefore it must be so that the integral involving the perpendicular part vanishes, so that

\begin{equation}
\iiint_{\mathbb{R}^3}\mathbf{\omega}\cdot ((\nabla \times \mathbf{\omega})\times \mathbf{u}_{\perp})dV=0
\end{equation}

This means that the enstrophy differential equation reduces to

 \begin{equation} \frac{1}{2}\frac{dE(t)}{dt}=\iiint_{\mathbb{R}^3}\mathbf{\omega}\cdot \nu \Delta \mathbf{\omega}dV\end{equation}
 
 and noting that the integral is always non-positive, finally we have
 
\begin{equation}  \frac{d}{dt}\int_{\mathbb{R}^3}\mathbf{\omega}\cdot \mathbf{\omega}d\mathbf{x}\leq 0\end{equation}

 It is well known that if  the total enstrophy stays bounded, the solutions stay regular.  QED.

\end{document}